\RequirePackage{fix-cm}
\documentclass[smallextended]{svjour3}       
\smartqed  
\usepackage{graphicx}
\usepackage{algorithm}
\usepackage{threeparttable}
\usepackage{multirow}
\usepackage{epstopdf}
\usepackage{subfig}
\usepackage{subfloat}
\usepackage{amsfonts}
\usepackage{mathrsfs}
\usepackage{array}
\usepackage{amssymb}
\usepackage{ulem}
\usepackage{bm}
\usepackage{amsopn}
\usepackage{amsmath}
\usepackage{color}
\usepackage{times}
\usepackage{fancyhdr}
\usepackage{algorithmic}
\usepackage{mathptmx}
\usepackage{cite}
\makeatletter
\renewcommand*{\@thmcounterend}{}
\makeatother

\spnewtheorem{lem}{Definition}{\bfseries}{\itshape}
\journalname{Machine Learning}

\begin{document}

\title{A Survey on Dynamic Network Embedding}


\author{Yu Xie         \and
        Chunyi Li \and
        Bin Yu  \and
        Chen Zhang \and
        Zhouhua Tang
}


\institute{ Yu Xie         \and
        Chunyi Li \and
        Bin Yu (Corresponding author) \and
        Chen Zhang \and
        Zhouhua Tang \at
              School of Computer Science and Technology, Xidian University, Xi'an, Shaanxi Province, China, 710071. \\
              Tel.: +029-88201901\\
              Fax: +029-88201901\\
              \email{yubin@mail.xidian.edu.cn}           
}

\date{Received: date / Accepted: date}

\maketitle

\begin{abstract}
Real-world networks are composed of diverse interacting and evolving entities, while most of existing researches simply characterize them as particular static networks, without consideration of the evolution trend in dynamic networks. Recently, significant progresses in tracking the properties of dynamic networks have been made, which exploit changes of entities and links in the network to devise network embedding techniques. Compared to widely proposed static network embedding methods, dynamic network embedding endeavors to encode nodes as low-dimensional dense representations that effectively preserve the network structures and the temporal dynamics, which is beneficial to multifarious downstream machine learning tasks. In this paper, we conduct a systematical survey on dynamic network embedding. In specific, basic concepts of dynamic network embedding are described, notably, we propose a novel taxonomy of existing dynamic network embedding techniques for the first time, including matrix factorization based, Skip-Gram based, auto-encoder based, neural networks based and other embedding methods. 
Additionally, we carefully summarize the commonly used datasets and a wide variety of subsequent tasks that dynamic network embedding can benefit. Afterwards and primarily, we suggest several challenges that the existing algorithms faced and outline possible directions to facilitate the future research, such as dynamic embedding models, large-scale dynamic networks, heterogeneous dynamic networks, dynamic attributed networks, task-oriented dynamic network embedding and more embedding spaces.

\end{abstract}

\section{Introduction}
\begin{figure}[!htb]
	\centering
	\tiny
	\subfloat{
		\includegraphics[width=0.45\linewidth]{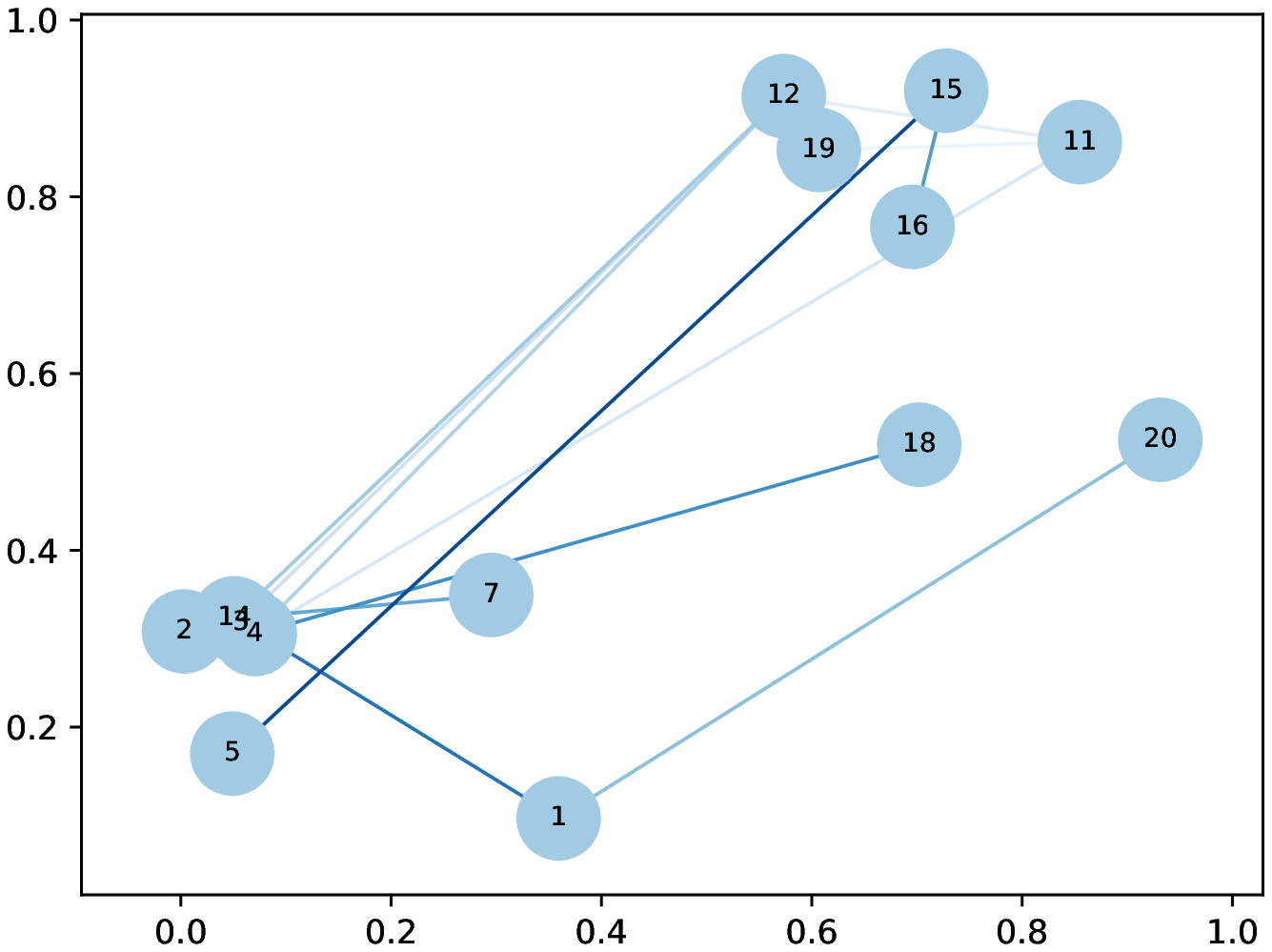}
		\includegraphics[width=0.45\linewidth]{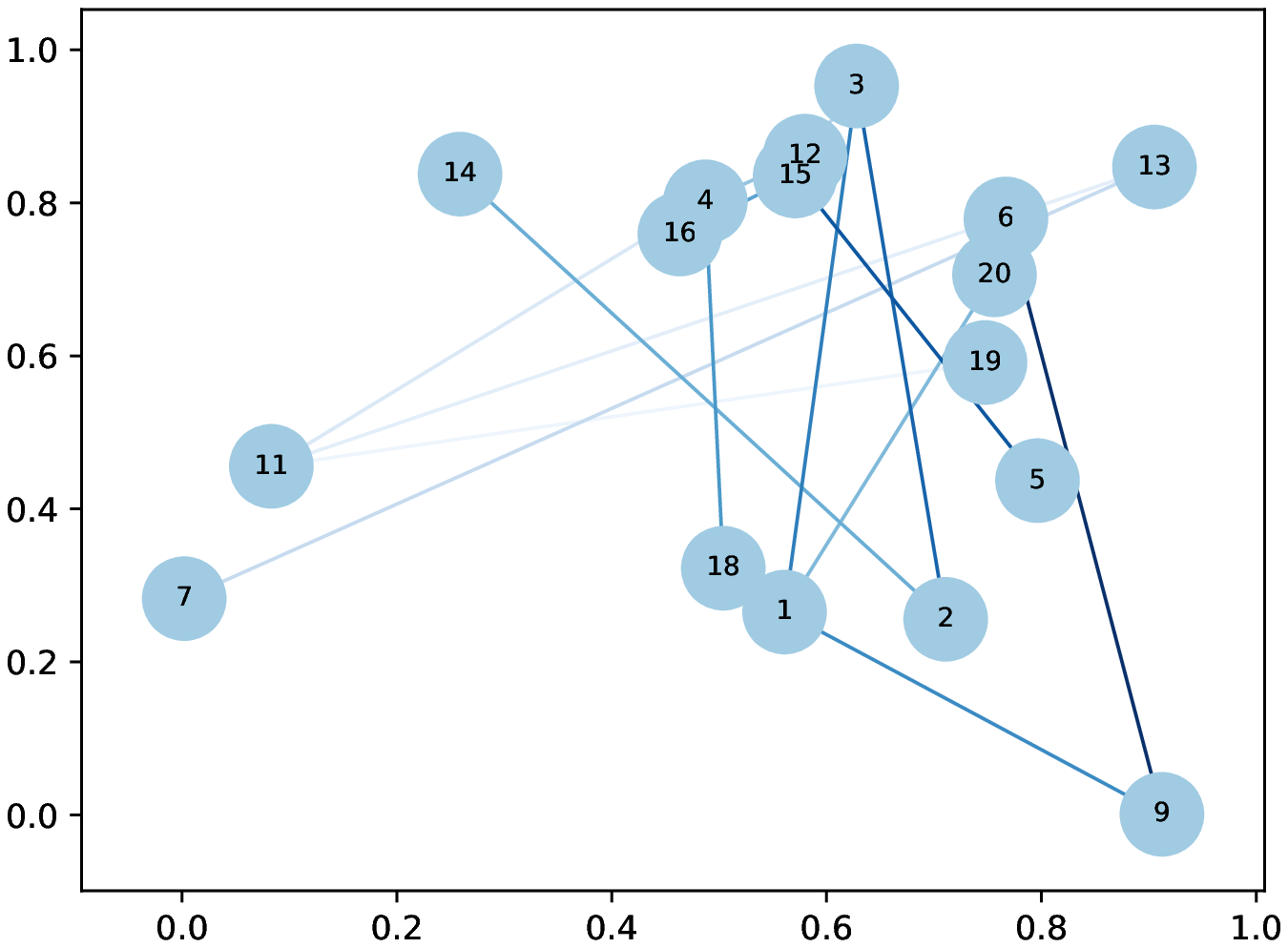}}
	\\
		\subfloat{
		\includegraphics[width=0.45\linewidth]{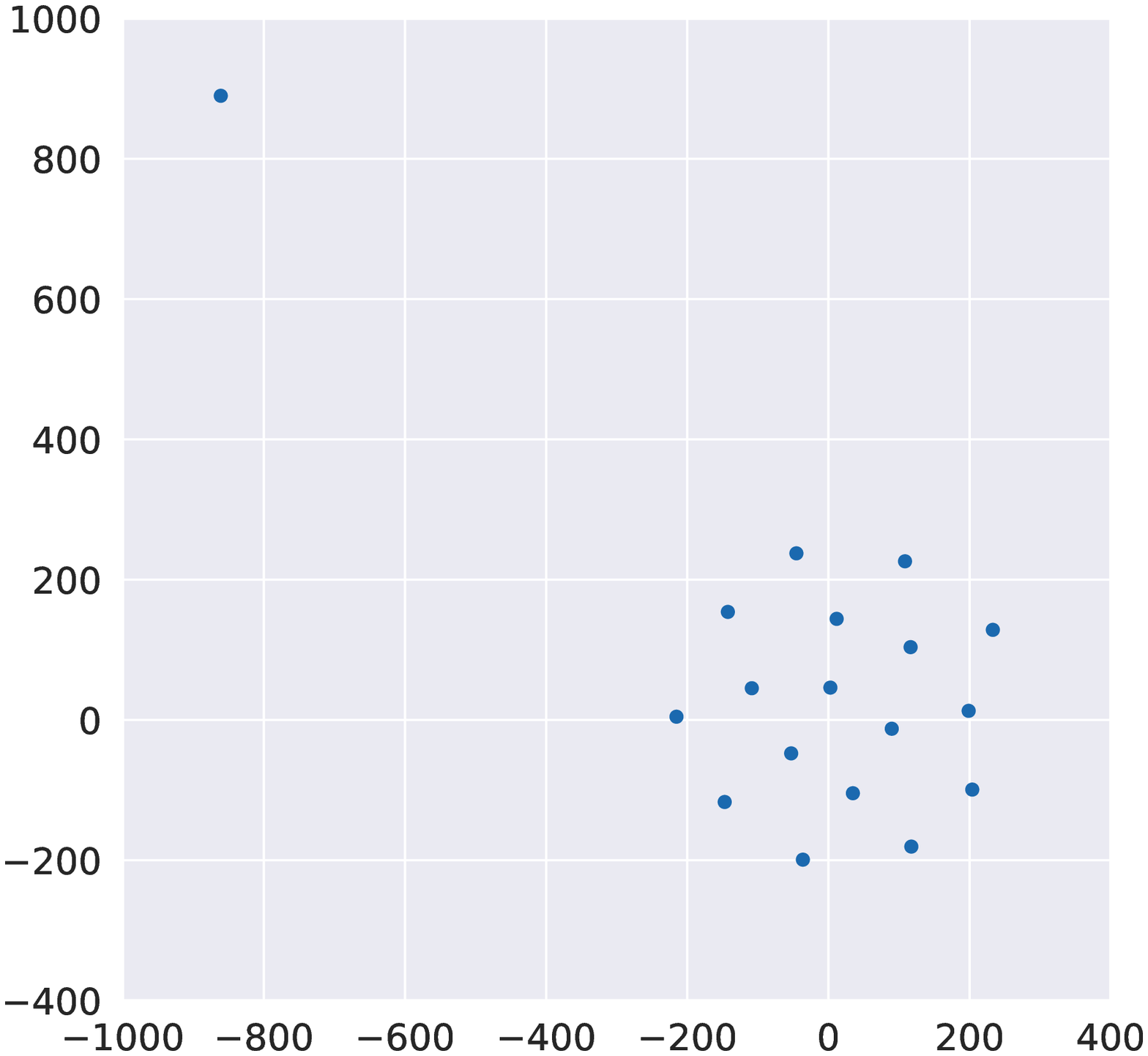}
		\includegraphics[width=0.45\linewidth]{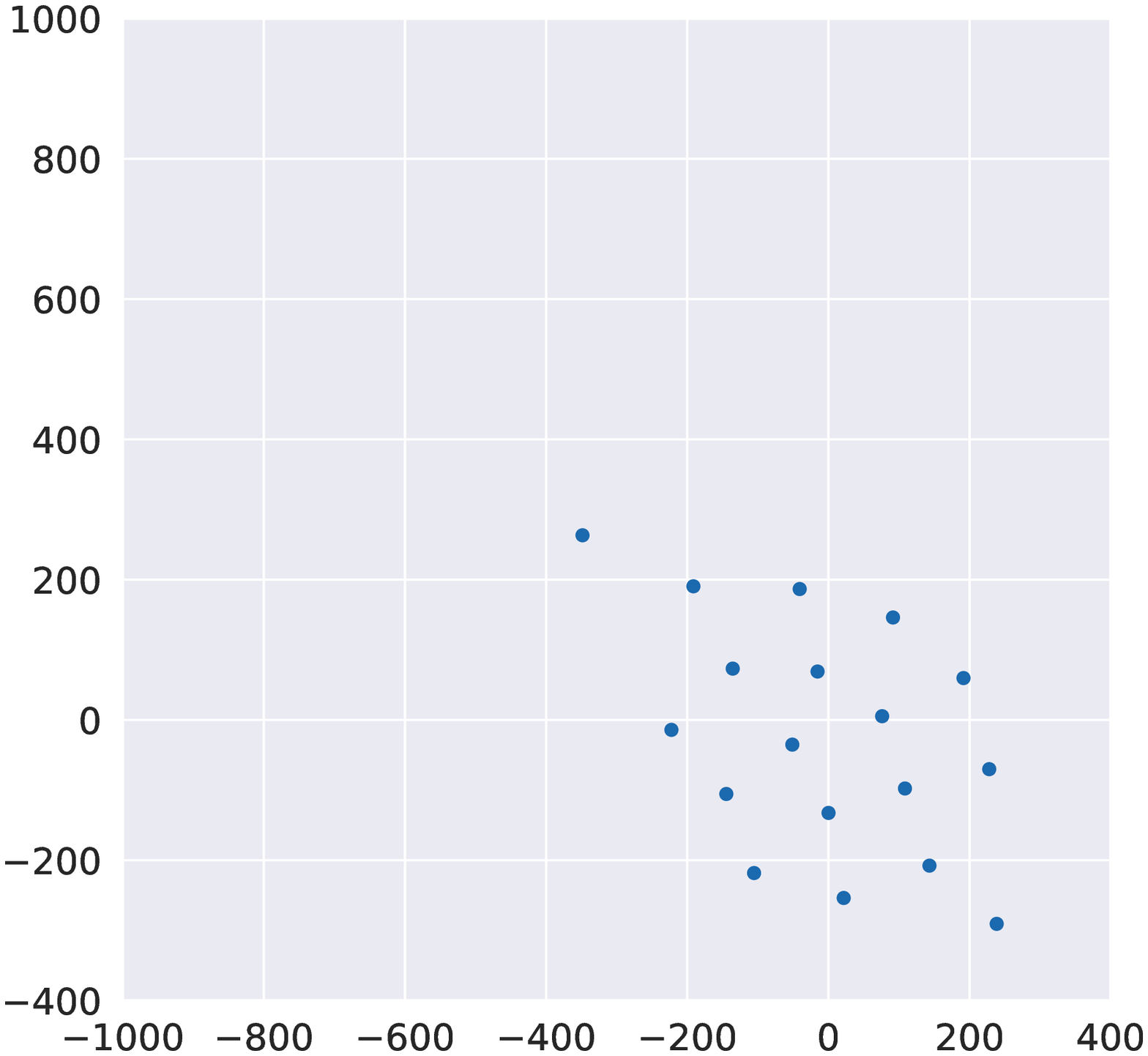}}
	\caption{Two snapshots of the datasets are collected in the MathOverflow social network in 2011-2012, an online community of interactive mathematics. For illustration, 20 nodes are selected randomly, and we draw the original network structure graph and its 2-D layout after embedding.}
	\label{figure1}
\end{figure}

Network analysis has attracted increasing attention over the recent years due to the ubiquity of network data in real world. The graph-structured network is a information carrier commonly used in complex systems, such as semantic networks \cite{yeon2020topical}, protein-protein interaction networks \cite{wang2019high}, social networks \cite{zhang2019relational} and criminal networks \cite{troncoso2020novel}. In order to construct the feature representations that can be applied to various tasks on graph-structured networks, network embedding is proposed to embed each node in the network into low-dimensional space \cite{cui2018survey}. However, the real-world networks are usually evolving, and simply exploiting static network embedding techniques for dynamic network embedding will increase the consumption of time and space, since the training model will be retrained once the network changes. Furthermore, in order to capture the temporal information in network evolution, many dynamic network embedding models are proposed to embed the unstructured data into a low-dimensional space, and predict the trend of network evolution. Fig. \ref{figure1} shows an illustrative example of embedding two snapshots of a dynamic network into a 2D space. Nowadays, dynamic network representation learning has been successfully applied to machine learning tasks on complex networks, such as visualization, node classification, node clustering, and link prediction.
Through modeling the interactions between entities, dynamic network embedding methods can facilitate practical applications, e.g., community detection \cite{hu2019efficient}, recommender system \cite{silveira2019good} and so on. Based on social networks whose interactions are constructed with the relationships, we can discover communities, build interpersonal networks, as well as predict interaction and behavior of users.

Considerable effort has been committed to develop static network embedding techniques. DeepWalk \cite{deepwalk:perozzi} formulates static network embedding as a sequence modeling problem, which generates node sequences by random walk and adopts Skip-Gram to predict the context from input nodes. Based on the DeepWalk model \cite{deepwalk:perozzi}, node2vec \cite{node2vec:grover} integrates depth-first search and breadth-first search strategies for exploring more flexible network structure. LINE \cite{line:tang} carefully designs a loss function for characterizing first-order and second-order proximities, and this method is applicable to large-scale information networks of arbitrary types. In addition, SDNE \cite{SDNE:wang} proposes a structural deep graph embedding model, which combines the advantages of first-order and second-order proximities, and maintains the highly non-linear structure. DNE-APP \cite{DBLP:conf/asunam/ShenC17} proposes a deep network embedding model, which adopts a semi-supervised stacked auto-encoder to obtain the compact representation. Furthermore, in order to prevent the manifold fracturing problem, ANE \cite{DBLP:conf/bigcomp/XiaoXHS18} utilizes an adversarial auto-encoder for variational inference to generate network representations by matching the posterior of node embeddings with an arbitrary prior distribution. Moreover, APNE \cite{2018apne} uses matrix decomposition to integrate structure and node label information. However, static networks lack the generalization ability in network evolution. For example, users can randomly create and delete their friendships on YouTube or Facebook, or active neurons tend to develop new relationships with other neurons in brain networks. The newly added or deleted edge has an impact on network analysis, while the existing static network embedding methods can not extract  rich temporal information. In order to overcome these obstacles, a great many dynamic network embedding methods are proposed.

In graph theory, a static network is composed of a set of nodes and a set of connected edges, which is indicated by an adjacency matrix. Dynamic networks are engaged with definitions which contain history information, snapshots, timestamps, nodes and edges. These definitions are concluded into two categories: snapshot and continuous-time networks. Snapshot is divided from the dynamic network at equal-interval in time sequence, so as to obtain the discrete sequence of network evolution through decomposing the dynamic network into a sequence of static networks. Nevertheless, dividing the network into equal-interval snapshots in time series places particular emphasis on global changes roughly and cannot retain the evolution information of the network elaborately. In contrast, the continuous-time mode can make up for this deficiency by marking each edge with multiple timestamps to predict whether the connection between nodes has just changed, in which each edge is given specific timestamps. Based on the two definitions mentioned above, several methods are positioned to make contributions to both research and practice in dynamic network embedding fields. In specific, Zhou \textit{et al.} \cite{DynamicTraid:zhou} designed a triadic unit, namely the dynamicTriad, which models how the triadic unit is closed from an open one, thus preserving both structural information and dynamics in network evolution. Nguen \textit{et al.} \cite{CDNE:Nguyen} incorporated temporal dependencies into node embedding and deep graph models, thereby integrating temporal information of dynamic networks. In \cite{dynnode2vec:mahdavi}, the evolving patterns in dynamic network are generated by using evolving random walks, and the current embedding vectors are initialized with previous embedding vectors. Ahmed \textit{et al.} \cite{Deepeye:ahmed} introduced a technology based on non-negative matrix factorization to obtain latent features from temporal snapshots of dynamic networks. Zhu \textit{et al.} \cite{DNPE:cui} proposed a generalized eigen perturbation model which can incorporate the variation of network links.

\begin{figure*}[!htb]
	\centering
	\includegraphics[width=1\linewidth]{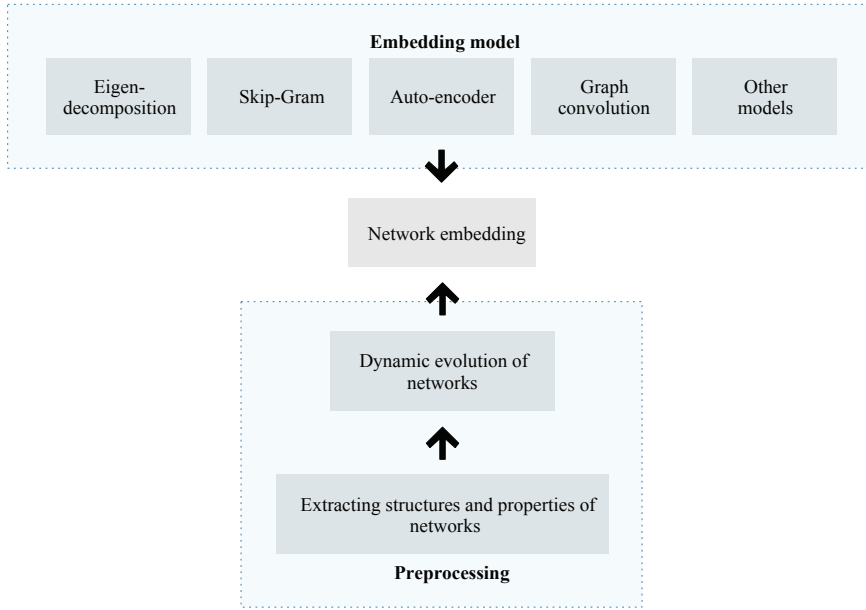}
	\caption{The content system of dynamic network embedding are constructed from two aspects, including embedding model and preprocessing process. In specific, the network dataset with attribute and label are necessarily preprocessed into uniform linear arrays, and then the normalized matrices are generated in the dynamic network evolution integrated with time series. After that, the network representations are constructed by updating the normalized matrices through the optimization algorithm.}
	\label{figure2}
\end{figure*}

According to both the attributes of the network and the architecture of the model, we categorize the existing dynamic network embedding techniques into five categories, including embedding based on eigenvalue factorization, embedding based on Skip-Gram, embedding based on auto-encoder, embedding based on graph convolution and embedding based on other methods. Fig. \ref{figure2} shows how to construct a system of dynamic network embedding by adopting a two-stage execution method. Different embedding methods have various criteria for assessing model performance and weaknesses. Theoretically, methods based on neural networks may suffer high computation and space cost.

Generalized from the classification scheme mentioned above, we introduce what challenges these categories of dynamic network embedding techniques confronted with, and how to address them. Although several surveys \cite{nishana2013graph}, \cite{Goyal11}, \cite{cui2018survey} summarize network embedding methods, they have two major limitations. On one hand, the existing graph embedding surveys only involve the relevant research based on traditional methods, and many recent innovative frameworks are not confirmed for inclusion in the review. For instance, \cite{Goyal11} mainly introduces three categories of approaches and lacks introductions to the novel methods. Moreover, \cite{WangMWG17} only shows solicitude for knowledge graph embedding. On the other hand, they only concern on static network embedding while ignoring the critical properties of dynamic networks. Despite remarkable advancements, the domain of dynamic network embedding still requires further research to make summaries on full-scale directions of novel techniques.

To summarize, we make the following contribution:
\begin{itemize}
	\item We propose a new taxonomy of dynamic network embedding according to different models and describe the challenges in these models, which open up new perspectives for understanding existing works. Furthermore, we systematically categorize the applications that dynamic network embedding supports.
	\item Our work follows the dynamic embedding approaches, but differs from previous works in listing the summarized insights on why dynamic graph embedding can be performed in a certain way, which can provide feasible guidance for future researches.
	\item To facilitate the timely and potential research, we point out six promising future research directions in dynamic network embedding domain, which consist of dynamic embedding models, large-scale networks, heterogeneous networks, attributed networks, task-oriented dynamic network embedding and more embedding spaces.
\end{itemize}

The rest of our paper is arranged as follows. In Section \ref{ProForm}, the problem definitions commonly used in the dynamic network embedding strategies are introduced, then we provide a formalized definition of dynamic network embedding. In Section \ref{Categories}, different dynamic network embedding techniques are elaborated in detail. The applications that dynamic network embedding enables are presented in Section 4. After that, Section 5 summarizes the challenges and points out several potential future research directions. In the end of this survey, we conclude the paper in Section 6.

\section{Problem Formalization}
\label{ProForm}

In Section \ref{Definition}, the definition of several basic concepts of dynamic network embedding are described in detail. In addition, Section \ref{Notations} provides the notations commonly used in this paper.

\subsection{Problem Definition}
\label{Definition}

\begin{lem}
	\textbf{(Network).} A network is usually represented as a graph $ G=\left ( V,E  \right ) $, where $ V $ denotes the node set and $ E $ denotes the edge set.
\end{lem}
\begin{lem}
	\textbf{(Static Network).} Given a network $ G=\left ( V,E  \right ) $, the network is static if all nodes and edges remain one state that are constant in time.
\end{lem}
\begin{lem}
	\textbf{(Dynamic Network).} A dynamic network $\mathcal{G}$ can be divided into a series of snapshots, i.e., $ \mathcal{G}=\left \{ G^{ 1 },...,G^{ T } \right \} $, where $ T $ denotes the number of snapshots. For each snapshot $ G^{ t } $, $ G^{ t }=\left ( V^{ t },E^{ t} \right ) $ indicates how nodes and edges are connected at the time step $ t $.
\end{lem}

\subsection{Notations}
\label{Notations}
The notations we used are defined in Table \ref{Tab 1}.

\renewcommand\arraystretch{1.5}
\begin{table}[!htbp]
	\setlength{\abovecaptionskip}{1pt}
	\caption{Notions and the descriptions}\label{Tab 1}
	\centering
	\begin{tabular}{|p{3.5cm}<{\centering}|p{4.2cm}<{\centering}|}
		\hline
		Notions & Descriptions \\ \hline
		$V$ & set of nodes \\ \hline
		$E$ & set of edges \\ \hline
		$d$ & embedding dimension \\ \hline
		$N$ & number of nodes \\ \hline
		$\mathcal{G}$ & a static network \\ \hline
		$ \mathcal{G}=\left \{ G^{ 1 },...,G^{T} \right \} $ & a dynamic network \\ \hline
		$ G^{T} $ &  a snapshot \\ \hline
		 $ Y^{T} $  &  embedding representations at time $T$ \\ \hline
		 $X$ & adjacency matrix \\ \hline
		 $\Delta X$ & change of adjacency matrix \\ \hline
		 $x_{1}, x_{2}, ... x_{n}$ & eigen vectors \\ \hline
		 $A$ & attributes matrix \\ \hline
		 $\Sigma$ &  diagonal matrix \\ \hline
		 $L$ & Laplacian matrix  \\ \hline
		 $S$ & collection of node pairs \\ \hline
	\end{tabular}
\end{table}

\section{Categories}
\label{Categories}
In this section, we introduce a new taxonomy for categorizing existing dynamic network embedding methods, as shown in Fig. \ref{figure2}. We adopt the classification strategies based on whether the dynamic network is represented as snapshots or continuous-time network marked by timestamps. According to the strategy, we categorize dynamic network embedding into five broad categories: embedding based on eigenvalue factorization, embedding based on Skip-Gram, embedding based on auto-encoder, embedding based on graph convolution and other embedding methods.
	
\subsection{Embedding Based on Matrix Factorization}
\label{Eigenvalue}
\begin{figure}[!htbp]
	\centering
	\includegraphics[width=1\linewidth]{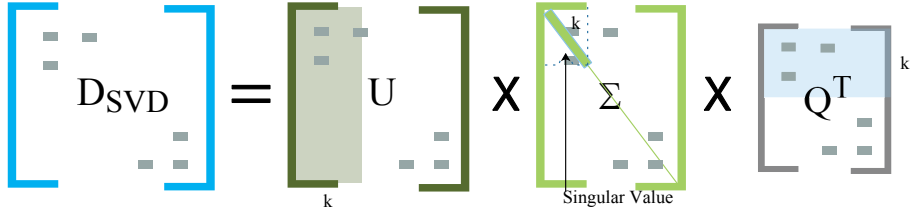}
	\caption{An illustration of singular value decomposition (SVD). $k$ is the rank of matrix $D_{\mathcal{SVD}}$.}
	\label{figure3}
\end{figure}

\begin{figure}[!htbp]
	\centering
	\includegraphics[width=1\linewidth]{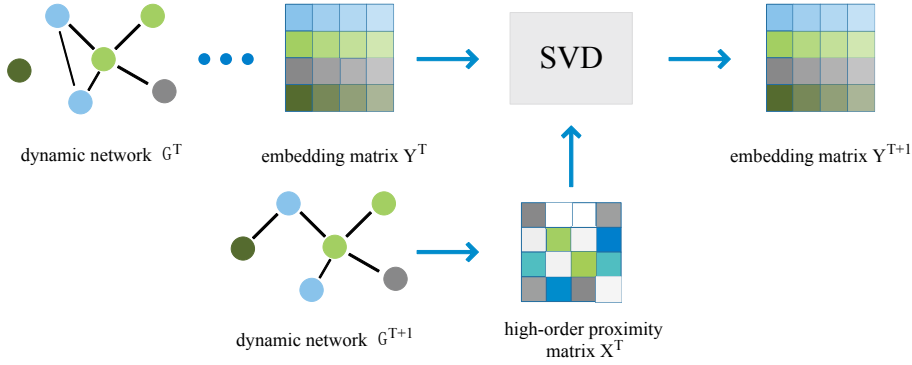}
	\caption{The model utilizes eigen decomposition to construct the high-order proximity matrix of the network, and then dynamically updates the node representations of the network in next time step based on the matrix perturbation theory.}
	\label{figure4}
\end{figure}

Matrix factorization \cite{Liu:2019} is the most general dimensionality reduction method in the field of network embedding. Singular value decomposition (SVD) is a representative algorithm of matrix factorization. The SVD of a matrix $D_{\mathcal{SVD}}\in R^{m \times n}$ is defined as:
\begin{equation}
D_{\mathcal{SVD}}=U \Sigma Q^{T}
\end{equation}
where $U \in R^{m \times m}$, $Q \in R^{n \times n}$, and $\Sigma \in R^{m \times n}$. $\Sigma$ is a diagonalized matrix, in which each element on the main diagonal is a singular value, all other elements are 0. Besides, matrices $U$ and $Q$ satisfy the equation $U^{T}U=I$ and $Q^{T}Q=I$. The  definition of SVD can be visually illustrated by the following Fig. \ref{figure3}.

A traditional but effective embedding method is to decompose the adjacency matrix by SVD and the attribute matrix by eigen decomposition in complex networks, so as to construct the embedding representation of each node. From the perspective of matrix factorization, the dynamic evolution of network is amount to the constant change of original matrix, where the network embedding is updated according to the perturbation theory of matrix. The overall perspective of eigenvalue factorization in dynamic network embedding is sketched in Fig. \ref{figure4}.
	
Li \textit{et al.} \cite{DANE:li} proposed a dynamic attributed network embedding model which represents dynamic networks as snapshots and considers situations in which adjacency and attribute matrices of the networks evolve over time. The dynamic network representation at the time $t$ is updated based on the matrix variation. For the time $t=1$, the model proposes an embedding algorithm combining the matrix $D$ and $X$ as the hot start. The formula for updating the embedded eigenvalues and eigenvectors is as follows:	
\begin{equation}
\begin{small}
\begin{array}{c}
{\Delta \lambda_{i}=x_{i}^{\prime} \Delta L_{X} x_{i}-\lambda_{i} x_{i}^{\prime} \Delta \Sigma_{X} x_{i}}
\\ {\!\Delta x_{i}\!=\!-\!\frac{1}{2} x_{i}^{\prime} \Delta_{X} x_{i} x_{i}\!+\!\sum_{j\!=\!2, j \neq i}^{k\!+\!1}\left(\frac{x_{j}^{\prime} \Delta L_{X} x_{i}\!-\!\lambda_{i} x_{j}^{\prime} \Delta \Sigma_{X} x_{i}}{\lambda_{i}-\lambda_{j}}\right) x_{j}}
\end{array}
\end{small}
\end{equation}
where $X^{t} \in \mathbb{R}^{n \times n}$ is the adjacency matrix of the attributed network at snapshot $t$ and $\Sigma_{X}$ is the diagonal matrix with $\mathrm{\Sigma}_{X}^{t}(i, i)= \sum_{j=1}^{n} X^{t}(i, j)$, $x_{1}, x_{2}, ... x_{n}$ are the eigenvectors of the corresponding eigenvalues, in the meantime $0=\lambda_{1} \leq \lambda_{2} \leq \ldots \leq \lambda_{n}$. $L_X^{t}=\Sigma_X^{t}-X^{t}$ , and $L_X$ is a Laplacian matrix. Moreover the network representation $Y^{ t+1 }$ is computed by using a correlation maximization method.	

\begin{figure*}[!htbp]
		\centering
		\includegraphics[width=1\linewidth]{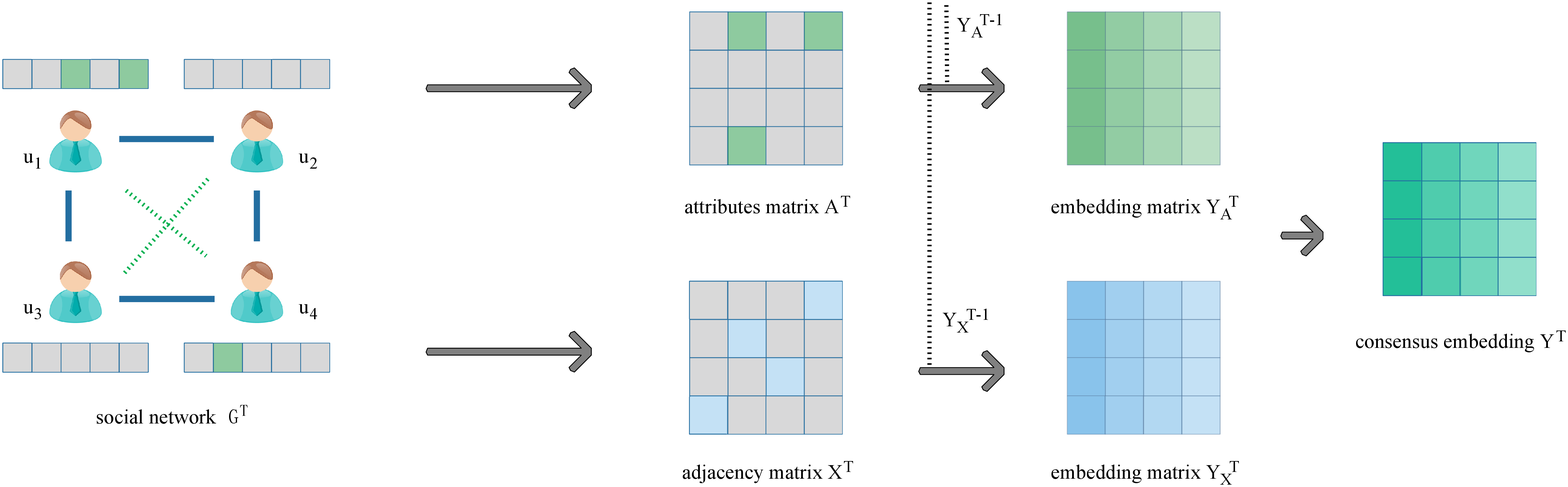}
		\label{dimension}
		\caption{An illustration of eigenvalue factorization in dynamic network embedding. At snapshot $t$, the model generates spectral embedding results by integrating with network structure $A$, and obtain the static embeddings $Y$. Afterwards, at the snapshot $t + 1$, topological change $\Delta A$ indicates the dynamic information of the network structure. The model integrates matrix perturbation theory for updating $Y$ for a new embedding $Y^{T}$.}
		\label{figure5}
\end{figure*}

Cui \textit{et al.} \cite{DNPE:cui} devised a embedding model which can capture the high-order proximity shown in Fig. \ref{figure5}, which is an extension of the asymmetric transitivity preserving graph embedding model\cite{ATPGE:ou} in the dynamic network processing. Based on the generalized singular value decomposition factorization (generalized singular value decomposition) and matrix perturbation theory \cite{theory}, the node representation of time $t+1$ is rapidly and effectively updated when the network structure changes in the next time step (adding/removing nodes/edges) while preserving the high-order proximity. Remarkably, Katz similarity is transformed into a generalized singular value decomposition factorization to model the high-dimensional structure of a dynamic network, and then the node representation at the time $t+1$ of the given network is dynamically updated by applying matrix perturbation theory.

Suppose $X$ denote the high-order adjacency matrix of the dynamic network, where $X^{K a t z}\!=\!M_{a}^{-1} M_{b}$. The Katz Index \cite{katz} is utilized as the alternate of $X$, which is the most generally used measures of high-order proximity. The formulas for updating the embedded eigenvalues and eigenvectors are shown as follows:
\begin{equation}
\begin{array}{c}{\Delta \lambda_{i}=\frac{x_{i}^{T} \Delta M_{b} x_{i}-\lambda_{i} x_{i}^{T} \Delta M_{a} x_{i}}{x_{i}^{T} M_{a} x_{i}}} \\ {\Delta x_{i}=\sum_{j=1, j \neq i}^{d} \alpha_{i j} x_{j}}\end{array}
\end{equation}
where $\left\{\lambda_{i}\right\}$ are the eigenvalues of $X$ in descending order and $x_{1}, x_{2}, ... x_{n}$ denote the corresponding eigenvectors, where d is the embedding dimension. As for $\alpha_{i j}$, it represents the coefficient matrix, which indicates the contribution of $x_{j}$ to $\Delta x_{i}$. Then the final embedded representation $Y$ is obtained by minimizing the distance between $X$ and $YY^{\mathsf{'T}}$.

Chen \textit{et al.} \cite{fast:chen} proposed two online algorithms to track the top eigen pairs in the dynamic network, which are capable of tracking the important network parameters determined by certain eigen functions. Besides, creative methods for analysis of eigen functions with attributions at each time stamp are proposed. Inspired by this, the eigen function can be utilized to map eigen pairs of the network to an attribute matrix:
	
\begin{equation}
	\mathrm{f} :\left(\Lambda_{k}, Y_{k}\right) \rightarrow \mathbb{R}^{x}(x \in \mathbb{N})
\end{equation}
where $\Lambda_{k}$ is the eigen pairs matrix transformed from the adjacency matrix $X$. $Y_{k}$ denotes the embedding matrix of the given network. After that, through the method of \cite{fast:chen}, the eigen function can be estimated by :
\begin{equation}
	f\left(\left(\Lambda_{k}, Y_{k}\right)\right)=\triangle(G)=\frac{1}{6} \sum_{i=1}^{k} \lambda_{i}^{3}
	\label{fast.1}
\end{equation}
where the quantity of triangles in the snapshot $G^{t}$ is $\triangle(G)$, and $\Lambda_{k}$ is the specific eigen-pair of $\Lambda_{k}$.
	
Ahmed \textit{et al.} \cite{Deepeye:ahmed} introduced a technology based on non-negative matrix factorization (NMF) to extract latent features which can strengthen the performance of the link prediction task on dynamic networks. This model focuses on the application of link prediction, and defines a new iterative rule to construct matrix factors with noteworthy network characteristics. In addition, it shows how to foster improvements in high prediction accuracy by adopting the fusion of time and structure information for training methods, and the potential NMF characteristics can effectively express the network dynamic rather than static representation.
	
However, approaches based on matrix factorization typically operates on more than one $n \times n$ matrices with a large dimension of $n \times n$, which suffers from high time complexity. To tackle this obstacle, Zhang \textit{et al.} \cite{TIMERS:zhang} proposed TIMERS to theoretically explore a lower bound of SVD minimum loss and replace the bound with the minimum loss on dynamic networks based on matrix perturbation. TIMERS optimally sets the restart time of SVD in order to reduce the error accumulation in time. Furthermore, driven by treating the maximum tolerated error as a threshold, TIMERS triggers SVD to restart automatically when the margin exceeds rated threshold. Besides, margin is a value between the reconstruction loss of incremental updates and the minimum loss.
	
\subsection{Embedding Based on Skip-Gram}
\label{Skip-Gram}
\begin{figure*}[!htbp]
	\centering
	\includegraphics[width=1\linewidth]{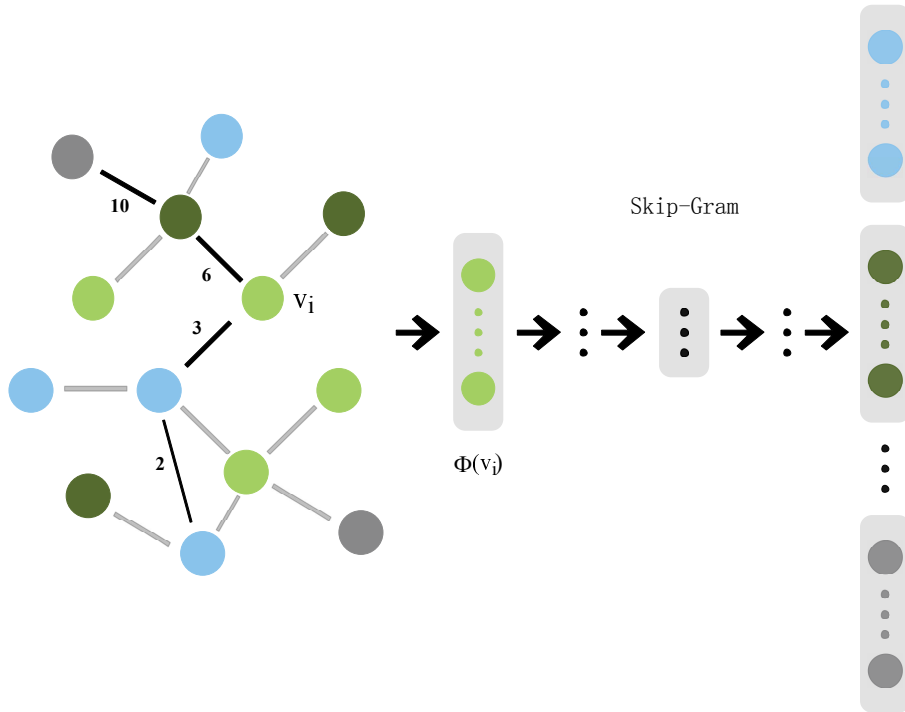}
	\caption{In the dynamic network, each edge is mapped to a corresponding timestamp. An effective walk is represented by a series of nodes connected by edges of an increasing timestamp. The mapping function $\Phi$ is a matrix, which maps every node into a $d$-dimensional vector with a total of $\left|V\right|\times d$ parameters which can be optimized by using Skip-Gram model or other models.}
	\label{figure6}
\end{figure*}

The great prevalence of Word2Vec \cite{word2vec:levy} has facilitated the well-known Skip-Gram model, through which we can predict the context from the input. Perozzi \textit{et al.} \cite{deepwalk:perozzi} suggested that nodes are converted into word vectors and a random walk sequence is viewed as sentences. Since then, a great deal of static network embedding algorithms are proposed based on this model. Fig. \ref{figure6} presents a synthesis framework which integrates temporal information into network embedding methods.
	
The following description outlines extension approaches. The classic node2vec \cite{node2vec:grover} inherits the advantages of DeepWalk and devises a biased random walk to learn node representations and LINE \cite{line:tang} can preserve the first-order and second-order proximities by utilizing multiple loss functions and optimize the performance by a concat technology. Du \textit{et al.} \cite{DNE:du} proposed a decomposable objective function on the basis of their previous LINE model to contribute the dynamic network embedding framework, where only a part of nodes are iteratively updated simultaneously and the representation in each iteration has a strong stability compared with retraining whole model. Moreover, the authors also devised the update node selection mechanism, greatly improving the efficiency of iterative update. Experiments on the node multi-label classification task \cite{Dyn2Vec:mitrovic} has performed well on several real networks. Similarly, Sajjad \textit{et al.} \cite{Sajjad:2019} proposed an efficient unsupervised network representation embedding method based on random walks that also divides networks into snapshots. In order to efficiently calculate the embedding results in the next snapshot which is impacted by the previous snapshot, this paper factorizes the process of generating network representations into two steps. Firstly, the authors proposed a random walk update algorithm responsible for updating the set of walk sequence, with the goal of statistically distinguishing the updated set of random walks from a set of random walks generated from scratch on the new graph. Secondly, the Skip-Gram model is modified to update the node representations based on the random walks.
	
Static network embedding usually obtains training corpus through random walks, and then integrates the corpus to the Skip-Gram model. However, the random walks fail to take the chronological order into account, in which condition the edges appear randomly. For instance, a message propagated in a network is directed, but an unconstrained random walk may result in a reverse corpus. At this point, Nguyen \textit{et al.} \cite{CDNE:Nguyen} expressed the dynamic network as a continuous-time network. Each edge has multiple timestamps, indicating the time corresponding to the existence of a connection relationship. On this basis, each random walk must be constrained to conform to the time sequence of occurrence of edges, so as to integrate the time sequence information of the network into the sequence of random walk. Theoretically, the random walk sequence with time series is a subset of which with non-temporal series. According to the view of information theory, the addition of temporal information reduces the uncertainty of random walk and enforces it outperforming DeepWalk and node2vec algorithms in traditional tasks.
	
Dynnode2vec \cite{dynnode2vec:mahdavi} modifies node2vec by regarding the prior embedding vectors as the initialization of a Skip-Gram model and employing the random walks in network volution to update the Skip-Gram based on previous timestamps information. Dynnode2vec modifies the static node2vec at time $t$ by capturing historical information from time step $t-1$. In dynnode2vec, only the evolving node embeddings are generated from sets of random walk sequences, rather than considering all nodes in the current timestamps. Hence, novel random walks from local changed regions can efficiently update the embedding vectors in network evolution.
	
Besides, dynamic network embedding framework proposed by Zuo \textit{et al.} \cite{HTNE:zuo} also integrates the Hawkes process based on temporal network embedding and the Skip-Gram models. This technology models the evolution process of nodes through neighbourhood formation sequence, and then captures the influence of historical neighbours on the current neighbourhood formation sequence by using Hawkes process.
	
Moreover, Liang \textit{et al.} \cite{twitter:liang} applied dynamic network embedding to user profiling in Twitter \cite{huang2020predicting}, and proposed a dynamic embedding model of user and word (DUWE), coupled with a model of streaming keyword diversification (SKDM). Particularly, DUWE aims at capturing the semantic information of users over time with the Skip-Gram model and attempts to maximize its log likelihood:
\begin{equation}
\log p\left(\mathrm{n}^{ \pm} | \mathrm{V}\right)=\sum_{k, l=1}^{V} n_{k, I}^{+} \log s\left(\mathrm{v}_{k}^{\mathrm{T}} \mathrm{v}_{l}\right)+n_{k, l}^{-} \log s\left(-\mathrm{v}_{k}^{\top} \mathrm{v}_{l}\right)
\end{equation}
where $n^{ \pm}=\left(n^{+}, n^{-}\right)$, $n^{+}, n^{-} \in \mathbb{R}^{V \times V}$ are the positive and negative indicator matrices for all word pairs with $n_{k, l}^{+}$ and $n_{k, l}^{-}$ being their elements respectively. $\left(v_{k}, v_{l}\right)$ is a word pair that can be observed in documents of a corpus. $s(x)$ is computed from a sigmoid function $s(x)=\frac{1}{1+\exp (-x)}$ and $s(-x)=1-s(x)$. $V=\left\{v_{k}\right\}_{k=1}^{V}$ is the definition of the embedding result of all the words in the vocabulary. DUWE is based on the Skip-Gram model, which is extended by employing a Kalman filter \cite{kalman:mihalcea} to process the evolution information of users and words, aiming to model the dynamic user and word embeddings. In DUWE, the authors utilized the variance of the transformation kernel embedded by all users to represent the diffusion process of the representation in users and words evolution.

\subsection{Embedding Based on Auto-encoder}
\label{Auto-encoder}
Auto-encoder is an artificial neural network that can construct the efficient representation of the input data in an unsupervised manner, which is suitable for dimensionality reduction. The dimension of the learned representation is generally far less than that of input data. In specific, auto-encoder is mainly carried out into two steps, encoder and decoder. In most auto-encoder, the hidden layers of the auto-encoder architecture is realized through the neural network.
\begin{figure*}[!t]
	\centering
	\includegraphics[width=1\linewidth]{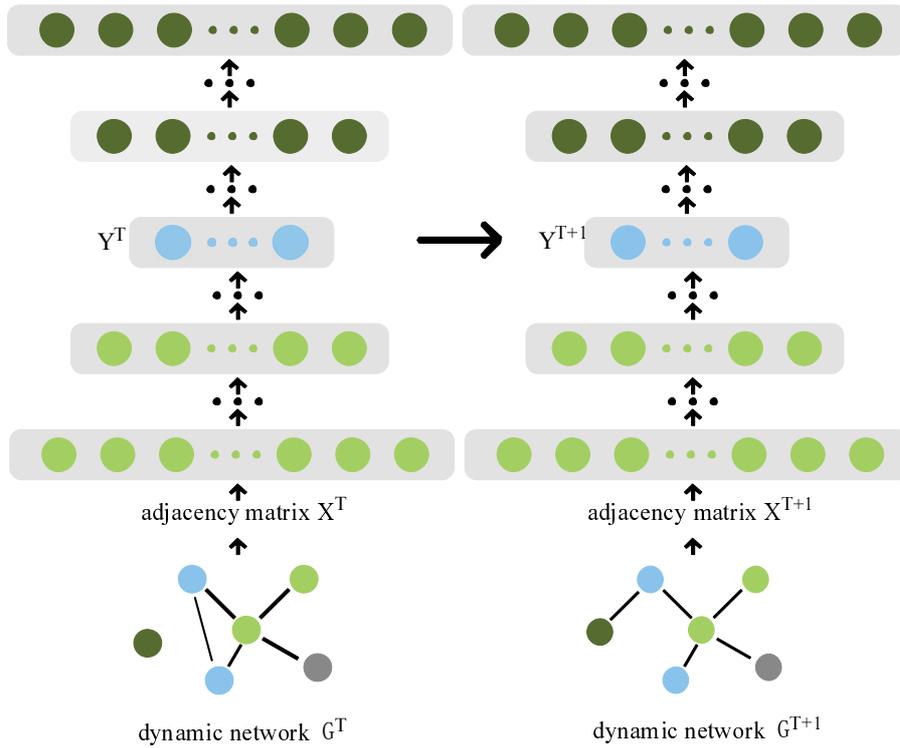}
	\caption{The dynamic network is divided into snapshots. During the training process, the weight parameters of the previous auto-encoder network are retained to initialize the next training network. $Y$ is the set of network representation, and $X$ is the adjacency matrix of networks.}
	\label{figure7}
\end{figure*}	

SDNE \cite{SDNE:wang} is a classical network embedding model based on auto-encoder. Due to its unsupervised nature and excellent performance, the embedding model can be extended to dynamic networks straightforwardly. For instance, Goyal \textit{et al.} \cite{GynGEM:Goyal} proposed a dynamic embedding model DynGEM, inspired by SDNE, which also represents dynamic networks as snapshots. It retains the embedding information at the previous moment that will be utilized for the next moment, so that the embedding model at the next moment can directly inherit the parameters of the model trained at the previous moment. It's worth noting that the quantity of network nodes and edges have an impact on the architecture of the embedding model. Therefore, this model employs heuristic information to adjust the overall structure of SDNE according to the new network structure, which makes it easy to be applied to the new network. A simplified version of the DynGEM model is shown in Fig. \ref{figure7}.
	
However, another work dyngraph2vec \cite{dyngraph2vec:Goyal} proposed by Goyal \textit{et al.} reflects that the DynGEM framework and other dynamic embedding algorithms capture the information of the previous step while ignoring the richer historical information. In this case, dyngraph2vec takes the network structure information into account at time step $t, t+1, t+2, \ldots \ldots, t+l-1$ when computing the embedding matrix at time $t+l$. To be specific, by expressing the time-ordered network sequence as a corpus, it leverages RNN to encode the historical information for semi-supervised learning and give a simple update rule for the auto-encoder. However, dyngraph2vec employs three models to embed the historical information, which leads to a high level of complexity. Additionally, the NetWalk model proposed by Yu \textit{et al.} \cite{NetWalk:yu} not only incrementally generates network representations in network evolution, but also detects network anomalies in real time. In particular, the node representation can be obtained through a number of walk sequences extracted from the dynamic network. It unifies local walks and hybridizes it with the hidden layer of a deep auto-encoder to yield node representations, thus, the resultant embedding is capable of reconstructing the original network with less loss. Moreover, the learning process performs over dynamic changes by utilizing a sampling stochastic. Then the model uses a dynamic clustering model to flag anomalous vertices or edges.
	
\subsection{Embedding Based on Neural Networks}
\label{Applications}
Recently, neural networks have attracted extensive attention of researchers in network embedding domain. By virtue of the powerful representation ability, neural networks based network embedding models have achieved outstanding results.
	
Since neural networks have shown impressive results on dynamic embedding, an inductive deep representation learning framework called DyRep \cite{RLDG:trivedi}, is decomposed into two dynamic processes on the network: association process and communication process. The former handles the change of network topology, while the latter deals with the change of network dynamics. In this model, event is defined to uniformly represent the changes of the above two processes. Furthermore, DyRep views the change of node representations as an intermediate bridge between the two processes mentioned above, so as to continuously update the node representation according to new events. When an event occurs, the new representation of the node is aggregated by the event-related neighbour nodes. In addition, the model which aims to aggregate neighbours of nodes dynamically adopts the attention mechanism.
	
After that, Chen \textit{et al.} \cite{scalable:chen} proposed a novel optimization method on neural networks, which assigns different sensitive weights for samples and selects the samples via their weights when computes gradients. The related samples are updated in a diffusion strategy, as the embedding of the selected sample is reconstructed. Meanwhile, in order to decrease the computational cost during selecting samples, the authors presented a nested segment tree for weighted sample selection. This model enforces dynamic embedding algorithms more adequate in analyzing highly dynamic and recency-sensitive data, which can overcome the problem that the optimization process of traditional is complicated.
	
Another example using neural network structure is Know-Evolve \cite{konw:trivedi}, it proposes a deep recurrent architecture for generating non-linearly network representations over time. It models multi-relational timestamped edges to reflect the evolving process. Besides, the information captured by node embeddings only depend on the edge-level information. Sankar \textit{et al.} \cite{DySAT:aravind} proposed DySAT, a dynamic self-attention network embedding model. Specifically, DySAT constructs node representations which incorporates self-attention mechanism into two aspects: structural neighbors and temporal dynamics. Through the dynamic self-attention architecture, it provides dynamic representations for nodes which capture both structural properties and temporal evolutionary patterns.
	
Recurrent neural networks (RNN) \cite{Lipton:2015} are commonly utilized for continuous sequence learning by capturing the dynamic information in time series. Motivated by the insights of RNN, several dynamic network embedding methods combine the RNN and other neural networks to learn the node representations. For example, EvolveGCN \cite{Pareja:2019} contains two components, the RNN component and the graph convolutional network (GCN) component. The RNN model updates the weight matrix in the next time step and the GCN model uses the updated weight matrix and the current node embedding matrix to update the subsequent node embedding matrix. EvolveGCN focuses on the evolution of the GCN parameters at every time step rather than the node representations, whose model is adaptive and flexible for preserving network dynamics.
	
In addition, BurstGraph \cite{BurstGraph} uses two RNN-based variational autoencoders framework to model graph evolution in the vanilla evolution and bursty links¡¯ occurrences at every time step. Furthermore, the encoders share the same GraphSAGE \cite{Hamilton:2017} to capture node attributed information. BurstGraph encodes both the structural evolution and attributed evolution in the latent embedding vectors, which can fully exploit the recurrent and consistent patterns in dynamic networks.
	
\subsection{Other Embedding Methods}
\label{Other}
Zhou \textit{et al.} presented DynamicTriad \cite{DynamicTraid:zhou}, which preserves the structural information of the network and incorporates the time-varying characteristics of the network into the embedding vectors. Through the analysis of triangular closure problems, the evolution of a dynamic network is devised to describe the change of the network. In order to express some rapidly changing nodes and reflect the evolution of the structure, the triangle closure problem is used to reflect network dynamics. In specific, for any three points, if they are connected by two nodes, they are called open triangles. In contrast, if they are connected by two nodes, they are called closed triangles. One of the characteristics of network structure evolution is that open triangle may evolve into closed triangle over time. By quantifying the probability of development from open triad to closed triad, the embedding vector of each vertex can be obtained at different time steps. Inspired by this ideas, DynamicTriad constructs the loss function to predict triangular changes:
	
\begin{equation}
\begin{small}
	L_{t r}^{t}=-\sum_{(i, j) \in S_{+}^{t}} \log P_{t r_{+}}^{t}(i, j)-\sum_{(i, j) \in S^{t}} \log P_{t r_{-}}^{t}(i, j)
\end{small}
\end{equation}	
where the first term is the probability of open triangle closing, and the second term is the probability that it is not closed. The representation $S_{+}^{t}$ is a collection of node pairs with no connection at the current time t, which will have edge connection at the next time. A collection of node pairs $S^{t}$ indicates that there will be no edge join at either the current time or the next time. Generally, the network is smooth, and the vector expression of the network varies little under the continuous timestamp. Therefore, combined with the rule smoothing term, the node representations can be trained.
	
Streaming graph neural networks (DGNN) proposed by Ma \textit{et al.} \cite{SGNN:yao} examines the relationship between the action time of edge connection and its effect on network embedding in a more detailed manner. It supposes that when new edges appear, both ends of the edges and the first-order neighbors of the endpoints have a high probability to make changes. However, the effect on the neighbors is related to the time difference. Neighbors who interact with the endpoints in the recent period are sensitive to new changes, while neighbors who interact over a relatively long period of time in the past are less affected. The endpoint and first-order neighbors are updated by two modules respectively and the temporal information enhancing long short-term memory model is adopted as the basic framework for updating modules to process information at different time intervals.

\section{Applications}
\label{Applications}
Network embedding can  improve the performance of various network mining tasks in both time and space. So this section gives a detailed description of several common network mining tasks that dynamic network embedding can benefit. We summarize the datasets and network mining tasks which are commonly used in proposed dynamic network embedding methods. Table \ref{Tab 2} concludes the statistics of each dataset used to perform different network mining tasks in dynamic network embedding experiments.
	
\renewcommand\arraystretch{1.5}
\begin{table*}[!htbp]
	\setlength{\abovecaptionskip}{1pt}
	\caption{A summary of datasets and network mining tasks}\label{Tab 2}
	\centering
\scriptsize
	\begin{tabular}{|c|c|c|c|c|}
		\hline
		Datasets & Node classification & Link prediction & Node clustering & Network visualization\\ \hline
		Academic & \checkmark & \checkmark & & \\ \hline
		Autonomous Systems(AS) &  & \checkmark &  &   \\ \hline
		CollegeMsg & & \checkmark &  &  \\ \hline
		DBLP & \checkmark & \checkmark & \checkmark & \checkmark \\ \hline
		Email-Eu-core&\checkmark  & \checkmark &\checkmark  &\checkmark   \\ \hline
		Enron &  & \checkmark &  & \checkmark \\ \hline
		Epinions & \checkmark & \checkmark & \checkmark & \checkmark  \\ \hline
		F2f-Resistance & \checkmark & \checkmark & \checkmark & \checkmark \\ \hline
		HEP-TH &  & \checkmark &  & \\ \hline
		Infectious &  & \checkmark &  &  \\ \hline
		Internet &  & \checkmark &  &  \\ \hline
		Loan & \checkmark & \checkmark & &  \\ \hline
		Mobile & \checkmark & \checkmark & & \\ \hline
		Synthetic Data (SYD) &  & \checkmark &  &  \\ \hline
		Twitter & \checkmark & \checkmark & \checkmark & \checkmark  \\ \hline
		Wikipedia &  & \checkmark & &  \\ \hline	
	\end{tabular}
\end{table*}
	
\subsection{Datasets}
This section describes the datasets commonly used in dynamic network embedding, which can be used for dynamic network analysis. For each dataset, specific preprocessing methods are adopted for different models. This paper lists the characteristics and detailed description of datasets commonly used in dynamic networks. Dynamic networks include synthetic dynamic graphs and real-world dynamic graphs. We summarize several commonly used dynamic datasets as follows.

\begin{itemize}
\item \textbf{Academic} \footnote{ http://www.aminer.org, an academic search engine\label{web}}.
It is an academic network composed of 51060 researchers as the vertices and 794552 coauthors as the edge. The size of snapshot is 2 years. In each snapshot, the existence of two users that have research cooperation projects will cause an edge created between two users. Each researcher has its specific community label.	

\item \textbf{Autonomous Systems(AS)} \cite{AS:2014}. The dataset is a communication network where each edge records the successful communication information between users and contains the communication records of 733 users from 1997 to 2000, which can be divided into snapshots by year.

\item \textbf{CollegeMsg} \footnote{ https://snap.stanford.edu/data/CollegeMsg.html\label{web}}.
The dataset consists of private messages sent on the University of California's online social network. Users initiate dialogs with other users in the network based on configuration file information. An edge $(u, v, t)$ indicates that user $u$ sends a message to user $v$ at time $t$. Within 193 days, it records the information of 1899 nodes and 59835 edges, which is divided into three snapshots.

\item \textbf{DBLP} \footnote{ http://snap.stanford.edu/data/com-DBLP.html\label{web}}. It is an academic website that counts the number of authors who have published more than two papers. From 2001 to 2016, the dataset restores information among 317080 authors and 1049866 edges, and each author own its research field label.

\item \textbf{Email-Eu-core} \footnote{ http://snap.stanford.edu/data/email-Eu-core-temporal.html\label{web}}.
The dataset is generated from e-mail network which is a large European research institute. It reflects anonymous communication of e-mail exchanges between members of the institute. There are 42 institutes in which all members are included. The e-mails only represent frequent daily communication between the core members of the organization. Moreover, it consists of 61046 e-mails and the time span is 803 days.
	
\item \textbf{ENRON} \cite{Enron:2004}. The dataset contains emails among employees in Enron Inc. from Jan 1999 to July 2002.

\item \textbf{Epinions} \footnote{ http://www.trustlet.org/wiki/Epinions\_datasets\label{web}}. Epinions is a goods review website where users post their comments and opinions about various goods. Besides, users themselves can build trust relationships, which is the basis for seeking advice between users. The dataset is a real-world dynamic network and the data has 16 different time stamps.

\item \textbf{F2f-Resistance} \footnote{ http://snap.stanford.edu/data/comm-f2f-Resistance.html\label{web}}.
This dataset is weighted, directed and temporal which contains networks extracted from 62 subgraphs, 451 nodes and 3,126,993 edges. Average temporal length per subgraphs 2,290 seconds which can be divided into 4 snapshots.
	
\item \textbf{HEP-TH} \cite{HEP-TH:2003}. This dataset records the collaboration among authors in academic conferences, which collected papers from the academic conference from January 1993 to April 2003.

\item \textbf{Infectious} \footnote{ http://konect.uni-koblenz.de/networks/sociopatterns-infectious\label{web}}. This network records the interaction of visitors in the exhibition. The formation of each edge requires at least 20 seconds of face-to-face communication between visitors with exact time stamp. It only stores visitors who communicate frequently.

\item \textbf{Internet} \footnote{ http://konect.uni-koblenz.de/networks/topology\label{web}}. This is the connection network of the internet autonomous systems. The nodes denote autonomous systems and edges represent the connections between autonomous systems. Multiple edges may connect two nodes, each of which represents an individual connection in time. Edges are	annotated with the time point of the connection.

\item \textbf{Loan}
The structure of this network is similar to that of the mobile dataset, which is provided by ppdai3, and contains 1603712 call records among 200000 registered ppdai users over 13 months. The size of snapshot is set to one month.

\item \textbf{Mobile}
It is a mobile network composed of more than 2 million call records provided by China Telecom. In 15 days, 340751 nodes and 2200203 edges were preserved. The user is regarded as the vertex and the snapshot is defined as a continuous one-day period without overlapping. In each snapshot, if one user calls another user, the connection between user vertices will be recorded in the network.

\item \textbf{Synthetic Data (SYD)}. Synthetic dynamic graphs are usually generated by the stochastic block model \cite{SBM:1987}. Snapshots of synthetic data are generated by setting the cross-block connectivity probability and the in-block connectivity probability.
	
\item \textbf{Twitter} \footnote{ https://bitbucket.org/sliang1/uct-dataset/get/UCT-Dataset.zip\label{web}}.
The dataset contains basic information of 1375 Twitter users, as well as blog information of all registered users as of May 31, 2015. The user's successful interaction information generated 3.78 million edges which have certain timestamp.

\item \textbf{Wikipedia} \footnote{ http://snap.stanford.edu/data/wiki-talk-temporal.html\label{web}}.
It is a dataset from Wikipedia, each edge represents the communication between users which has a specific time label. The time span is 2320 days, including 1140149 registered users and 7833140 connections.

\end{itemize}
	
\subsection{Node Classification}
Node classification aims to classify each node into corresponding categories and facilitate the downstream applications, such as recommendation, targeted advertising, interest mining and interpersonal network building. The classification technique is used to label the nodes in the dynamic network, which is conducive to the in-depth analysis of the characteristics of the network structure and the extended study of the application. Node classification includes multi-label and multi-class classifications. The former can be conducted on datasets where each node is only labeled by one category. The latter is conducted on datasets where each node is labeled by more than one category. The process of node classification adopted for network embedding is generally decomposed into four steps. Firstly, a network embedding algorithm is applied to derive the embedding representations of networks. Secondly, the dataset with label information needs to be divided into training and testing sets. Then, train a classifier on the training set. Lastly, perform node classification on the testing set. The selectable classifiers contain Liblinear \cite{Liblinear:2008}, nearest neighbor classifier \cite{NearestNeighborClassifier:2011}, support vector machine(SVM) \cite{SVM:1999} and so on. Micro-F1 and Macro-F1 are commonly utilized as the evaluation metrics for the node classification task which integrate both the accuracy rate and recall rate of the classification model and are defined as follows:

\begin{equation}
Micro-f1= \frac{{2 \times P \times R}}{{ P+R}}
\end{equation}

\begin{equation}
Macro-f1= \frac{{\sum\nolimits_{M \in C } {F1(M)}}}{{\left|C\right| }}
\end{equation}
where $F1(M)$ is the $F1$-measure of label $M$, and $C$ is the whole label set. Moreover, $P$ indicates the precision rate and $R$ indicates the recall rate.

\subsection{Link Prediction}
The link prediction task \cite{bu2019link} is target to predict whether there exists an edge between two nodes, which can evaluate the performance of embedding methods on preserving explicit network topological structures. Meanwhile, it is also solved by using a binary classification technology, in which the input is the learned embedding representations of nodes and the output is a the predicted result of a binary classifier. If there exists an edge between the two input nodes, the output is $ 1 $, otherwise, the output is $ 0 $. Link prediction adopted for dynamic network embedding is divided into three steps. First of all, learn embedding representations of nodes. Then, divide the dataset into a training set and a testing set. For any pair of nodes in the training and testing sets,  a label is generated to record whether these two nodes have connections. Finally, conduct link prediction experiments on the testing set. The AUC \cite{AUC:2006} and MAP \cite{MAP:2008} measures are employed as the evaluation metric. MAP aims at measuring models with recall and precision which can reflect the global performance. Compared with AUC, MAP considers the performance ranking of returned items. The formula of MAP is

\begin{equation}
MAP = \frac{{\sum\nolimits_1^n {AP(i)} }}{{|V|}}
\label{eq17}
\end{equation}
\begin{equation}
AP(i) = \frac{{\sum\nolimits_j {precision@j(i) \cdot {\Delta _i}(j)} }}{{|\{ {\Delta _i}(j) = 1\} |}}
\label{eq18}
\end{equation}
where $precision@j(i)$ is the precision of node $v_{i}$. $\Delta {}_i(j)=1$ presents a connection between nodes $i$ and $j$.

\subsection{Node Clustering}
As an unsupervised task, the objective of node clustering is dividing a network into several clusters, where nodes belonging to the same cluster have more similarity. Analogous to node classification, node clustering endeavors to learn the category information of nodes. After node representations are obtained from embedding algorithms, some typical clustering methods, for example Kmeans \cite{kmeans:1967}, are adopted actually for performing node clustering. The normalized mutual information (NMI) \cite{NMI:2011} is commonly utilized to evaluate the clustering performance of network embedding algorithms. The function is shown as follows:
\begin{equation}
\begin{tiny}
NMI\!=\!1\!-\!\frac{H(M_1|M_2)_{norm}\!+\!H(M_2|M_1)_{norm}}{2}
\end{tiny}
\end{equation}
where $NMI$ is used to measure the similarity between $M_1$ and $M_2$ clustering results. $H(\cdot)$ is the formula of information entropy. To evaluate a model, if nodes that belong to the same category are clustered into the same cluster based on obtained node representations, the embedding algorithm learn desirable node representations.
	
\subsection{Network Visualization}
Network visualization is also a widespread application in dynamic network embedding. In order to generate meaningful visualization results, a network is mapped into two-dimensional space. Embedding representations of nodes are used as the input of visualization tools, such as t-SNE \cite{t-SNE:2008}. After node representations are generated by embedding algorithms, t-SNE first maps the $ d $-dimensional node representations into a 2D space which the visualization results are displayed in. Network visualization is conducted on datasets with labels. Nodes with different labels are marked with different colors in the 2D space. Because network embedding preserves the intrinsic structures of a network, visualization results directly reflect that the nodes in the same cluster in two-dimensional space have more similar characteristics.
	
\section{Future Research Directions}
The mentioned survey on the traditional and novel dynamic network embedding methods demonstrates that putting forward creative methods in dynamic networks to effectively improve the performance of downstream network analysis tasks is still a necessary task. Because networks are evolving over time and the above models are only limited to two mode, snapshot and continuous-time mode, which are difficult to adapt for more complex situation. Therefore, we discuss several promising research directions for future works in this section.

\subsection{Dynamic Embedding Models}
In this survey, we roughly divide the existing dynamic network embedding methods into five types of models, i.e., matrix factorization based models, Skip-Gram based models, auto-encoder based models, graph convolution based models and other models. However, all of these models are inspired by static embedding methods. As shown in Fig. \ref{figure2}, for all existing dynamic embedding methods, the dynamic is reflected in the process of handling the original network rather than the process of embedding the network. In fact, there are few works that implements network embedding execution with considering real-time network evolution. For example, if some edges are added or removed in a network, existing methods need to reprocess the original network and then perform network embedding, which is time-consuming. Aiming at effectively updating the embedding representations when the status of a network changes, dynamic embedding models are highly desirable.
	
\subsection{Dynamic Network Embedding for Large-scale Networks}
For network embedding, Tang \textit{et al.} \cite{line:tang} proposed LINE to analyze complex networks with large-scale. However, LINE is not applicable for dynamic networks. Due to the complexity of performing network evolution in dynamic networks, existing dynamic network embedding methods can not adopt for more complex real-world networks. By the insight of the above survey, dynamic networks can be represented as snapshots or continuous-time networks marked by timestamps. The more the number of snapshots or timestamps, the higher complexity of network evolution. Therefore, the efficiency of dynamic network embedding can be improved from two aspects, i.e., reducing the complexity of network evolution or improving embedding models. To conclude, large-scale dynamic network embedding is a difficult research point without effective work reaching it.
	
\subsection{Dynamic Network Embedding on Heterogeneous Networks}
Although many researchers focus on dynamic network embedding, existing methods are mainly proposed based on homogeneous networks that contain single type of nodes and edges. Nevertheless, it is well recognized that many real-world networks are heterogeneous networks where several types of nodes and edges exist. Dynamic embedding of heterogeneous networks need to consider node types and edge types when networks change over time. What's more, if the structures and properties are extracted from the original heterogeneous networks, the extracted structures and properties should be preserved as much as possible in the process of network evolution. Exploring the domain of dynamic network embedding on heterogeneous networks is an interesting research direction.
	
\subsection{Dynamic Network Embedding on Attributed Networks}
Attributed networks contain extra attributed information, such as texts or contents \cite{yu2020structured}. Dynamic embedding of attributed networks emphasizes the information that both the structures of networks and the attributes of nodes provide as networks evolve. A little work about this research direction has been proposed. Only Li \textit{et al.} \cite{DANE:li} proposed DANE that updates both the adjacency and attribute matrices when a network changes over time. Nevertheless, the work of Li \textit{et al.} \cite{DANE:li} only researches the situation where a dynamic network is divided into sequential snapshots. There is no research to pay attention to the other situation in which the dynamic network is represented as continuous-time networks marked by timestamps.

\subsection{Task-oriented Dynamic Network Embedding}
As mentioned above, network mining tasks include link prediction, node classification, node clustering, network visualization and so on. Network embedding methods oriented by different network mining tasks are proposed for static networks. For example, Yang \textit{et al.} \cite{Yang:2014} described a network embedding technology for node classification. SHINE \cite{SHINE:2018} presents a signed heterogeneous information network embedding framework for sentiment link prediction. Compared with universal network embedding algorithms that can be utilized to conduct different network mining tasks, task-oriented network embedding methods focus on only one task so that some extra information related to the task can be extracted to train the embedding models. Task-oriented embedding models are more effective on the target task than universal embedding models. However, there is no related work to address a single network mining task in dynamic network embedding.
	
\subsection{More Embedding Spaces}
In the previous works, EOE \cite{EOE:2017} presents a joint deep embedding framework for coupled heterogeneous networks, which embeds a network in two embedding spaces. However, EOE is designed for static networks. According to the existing methods mentioned above, the dynamic network is commonly represented in one embedding space. In general, the deeper embedding with more target spaces means the higher complexity of time and space. But on the other hand, embedding a network into more spaces mines more possibilities of preserving network structures and networks. Therefore, exploring different embedding spaces is an interesting and challenging research direction for dynamic network embedding.
	
\section{Conclusion}
In this paper, we summarize the existing dynamic network embedding methods. A dynamic network can be divided into sequential snapshots or continuous-time networks marked by timestamps. Based on this premise, we divide dynamic network embedding models into five categories: embedding models based on eigenvalue decomposition, embedding models based on Skip-Gram, embedding models based on autoencoders,	embedding models based on graph convolution and other embedding models. Almost all existing dynamic network embedding methods are generalized into these five categories. For each category, we make a detailed description. What's more, the applications of dynamic network embedding are also included in our survey. In the section of applications, the datasets and network mining tasks that are widely adopted in dynamic network embedding are described in detail. Afterwards and primarily, we provide six interesting and promising future research directions.

\section*{Acknowledgement}
The authors wish to thank the editors and anonymous reviewers for their valuable comments and helpful suggestions which greatly improved the paper's quality. This work was supported by the Key Research and Development Program of Shaanxi Province (Grant no. 2019ZDLGY17-01, 2019GY-042).


%
%



\bibliographystyle{spmpsci}
\bibliography{mybibfile}

\end{document}